\documentclass[%
 reprint,
 amsmath,amssymb,
 aps,
]{revtex4-2}
\usepackage[whole]{bxcjkjatype}
\usepackage{graphicx}
\usepackage{dcolumn}
\usepackage{bm}
\usepackage{braket}

\begin{document}

\preprint{APS/123-QED}

\title{Enhancement of femtosecond photon echo signals from an inhomogeneously broadened InAs quantum dot ensemble using chirped pulses}

\author{Yuta Kochi}
\email{yuta.kouchi@keio.jp}
\author{Yutaro Kinoshita}
\author{Masanari Watanabe}
\author{Ryutaro Ide}
\author{Junko Ishi-Hayase}
\email{hayase@appi.keio.ac.jp}
\affiliation{Keio University}
\affiliation{Center for Spintronics Research Network, Keio University}

\author{Kouichi Akahane}
\affiliation{National Institute of Information and Communications Technology}

\date{\today}

\begin{abstract}
Photon echo (PE) techniques offer a promising approach to optical quantum memory, yet their implementation in conventional platforms, such as rare-earth-ion-doped crystals, is hindered by limited bandwidths. Semiconductor quantum dot (QD) ensembles, featuring THz-scale inhomogeneous broadening and sub-picosecond dynamics, provide an attractive alternative for ultrafast applications. However, achieving coherent control across such broad spectral ranges remains challenging due to detuning and spatial field inhomogeneities, which reduce PE efficiency. In this work, we experimentally demonstrated adiabatic rapid passage (ARP)-enhanced PE in dense, self-assembled InAs QD ensembles exhibiting THz-scale inhomogeneous broadening and operating at telecom wavelengths, achieving a 3.2-fold increase in echo efficiency. Chirped control pulses designed to satisfy adiabatic conditions across the ensemble enable broadband rephasing. Numerical simulations based on a two-level model reproduce the key experimental observations, including the ARP-induced enhancement, thereby validating the underlying physical picture. These results establish ARP as a robust and scalable approach for coherent control in InAs QD ensembles, with potential applications for ultrafast and broadband optical communication in the THz spectral region.
\end{abstract}

\maketitle


\section{\label{sec:level1}Introduction}
Photon echo (PE) techniques have long served as powerful tools for probing coherence properties and phase relaxation dynamics in inhomogeneously broadened media, offering valuable insights into quantum information science~\cite{Sangouard11}. PE can be understood as a specific form of four-wave mixing (FWM), a third-order nonlinear optical process governed by the material’s $\chi^{(3)}$ susceptibility. In this process, macroscopic coherence induced by an initial excitation pulse rapidly dephases due to the inhomogeneous distribution of resonance frequencies across the ensemble. A subsequent rephasing pulse recovers this dephasing, resulting in the emission of a coherent echo at a later time. This rephasing effectively compensates for inhomogeneous broadening and recalls ensemble coherence, providing a direct probe of collective quantum dynamics.

In recent years, PE-based protocols have attracted renewed attention as promising candidates for optical quantum memory, owing to their intrinsic multimode capacity and compatibility with time-domain multiplexing~\cite{McAuslan11, Pascual13}. Among various implementations, rare-earth-ion–doped systems such as $\mathrm{Eu}^{3+}$:$\mathrm{Y}_2\mathrm{SiO}_5$ crystals and erbium-doped fibers have emerged as leading platforms
~\cite{Afzelius09, Afzelius10, Damon_2011, Bonarota_2014, Beavan:11, Holzapfel_2020, Ortu22, Wei24}. However, their operational bandwidths—typically on the order of GHz—fundamentally limit their suitability for storing sub-nanosecond optical pulses.

Semiconductor quantum dot (QD) ensembles have attracted significant interest as alternative platforms due to their THz-scale inhomogeneous broadening and sub-picosecond excitonic dynamics~\cite{Borri01, Kamada01, Ikezawa07, Poltavtsev16, Mishra17, Wei20, Kosarev20, Kosarev22, Heindel23, Nikoghosyan23, Trifonov24}. In particular, self-assembled InAs QDs operating at telecom wavelengths offer compatibility with existing fiber-optic infrastructure and enable direct manipulation using femtosecond optical pulses~\cite{Hayase07}. Nevertheless, coherent control of such broadband and spatially inhomogeneous ensembles remains challenging. Strong spectral detuning and spatial variation in the excitation field amplitude often result in incomplete rephasing and reduced PE efficiency.

To overcome these limitations, adiabatic rapid passage (ARP) has been explored as a robust technique for quantum state control. ARP employs chirped pulses to achieve population inversion even in the presence of detuning and field inhomogeneities~\cite{VITANOV200155, Gawarecki12, Gabor13, Li23}. It has been successfully employed to enhance electromagnetically induced transparency (EIT) in $\mathrm{Pr}^{3+}$:$\mathrm{Y}_2\mathrm{SiO}_5$~\cite{Mieth12}, and to control single or few QDs in nanophotonic structures~\cite{Wu11, Wilbur22, Kappe24, Simon11, Mukherjee20, Schuh11, Ramachandran21, Ramachandran24}. Moreover, ARP has also been investigated in nonlinear optical processes, including FWM~\cite{Eyal18, Xiaoyue20}, and has been demonstrated to improve PE efficiency in rare-earth-ion–doped crystal quantum memories~\cite{Damon_2011, O'Sullivan22}. However, in QD-based studies, ARP has primarily been employed for reliable population inversion at the single-emitter level, with limited attention to dense QD ensemble coherence. Thus, the potential of ARP for coherent control in dense QD ensembles with THz-scale inhomogeneous broadening remains largely unexplored.

In this work, we report the first demonstration of ARP-enhanced photon echo generation in a self-assembled InAs QD ensemble operating in the telecom band. By employing chirped optical pulses designed to satisfy ARP conditions across the THz-scale inhomogeneous broadening—approximately three orders of magnitude wider than in rare-earth systems—we achieve a pronounced enhancement of ensemble coherence. This enhancement is evidenced by stronger PE signals, despite the need for large chirp bandwidths and robust adiabatic tuning. Numerical simulations based on a two-level model with realistic system parameters reproduce the observed dynamics and elucidate the underlying mechanisms. Our results establish ARP as a scalable and robust method for broadband coherent control of ultrafast nonlinear processes such as photon echo and four-wave mixing. Extending ARP into the femtosecond regime enables manipulation of complex many-body coherences on unprecedented timescales, opening promising pathways toward ultrafast quantum memory devices and broadband photonic quantum technologies compatible with existing telecom infrastructure.

\section{\label{sec:level1}Theoretical Modeling and Experimental Setup}
\subsection{\label{sec:level2}Single pulse ARP dynamics in a two-level system}

\begin{figure}[t]
\includegraphics[width=\columnwidth]{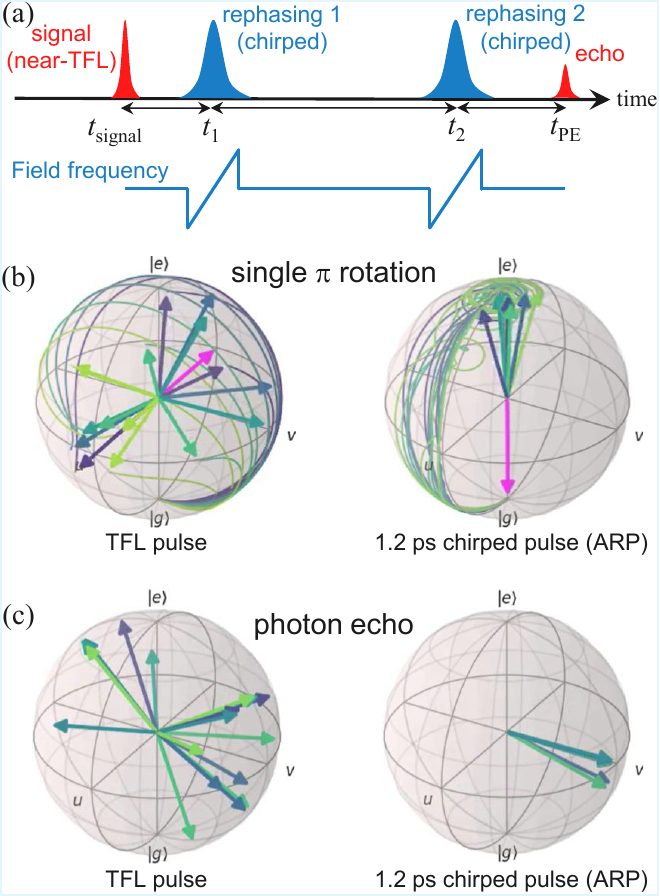}
\caption{(a) PE pulse sequence of our setup. We applied double rephasing pulses (rephasing 1 and 2). Blue line represents the change in frequency of the rephasing pulses. (b) Comparison of final Bloch vector states in population inversion for single TFL and 1.2 ps chirped pulse (ARP) excitation. The magenta arrows represent the final positions of the torque vectors, while the solid lines trace the trajectories of the Bloch vectors. (c) Comparison of final Bloch vector states in PE for TFL and ARP rephasings, starting from the superposition state (v in this figure) as the initial condition.}
\end{figure}

\begin{figure}
\includegraphics[width=\linewidth]{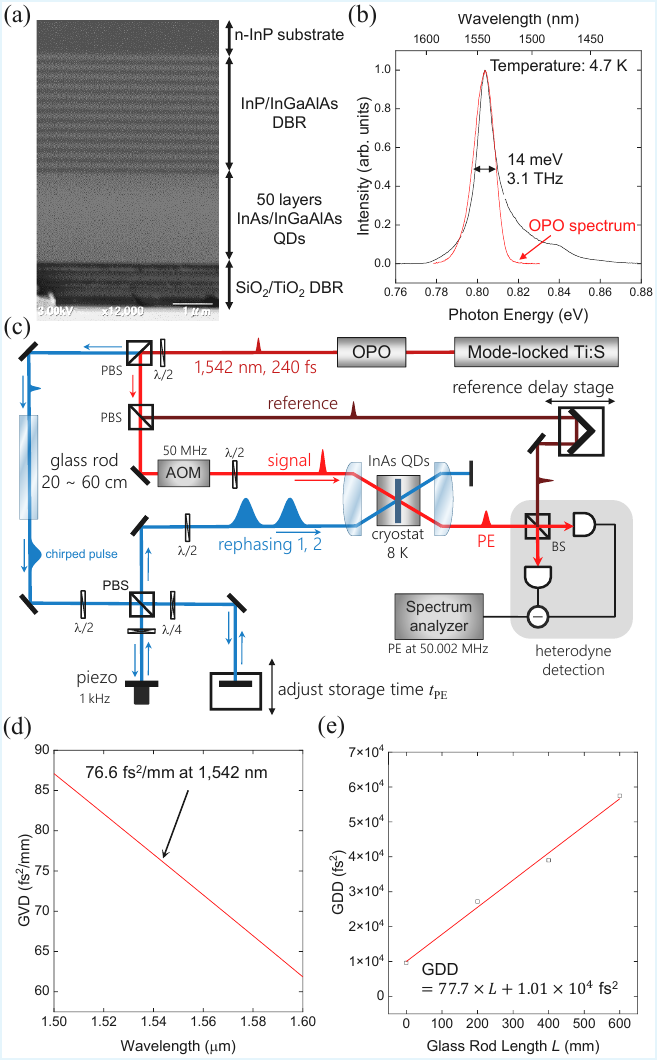}
\caption{(a) Structure of the InAs QD ensemble integrated with a resonator. (b) Photoluminescence spectrum of the QD ensemble at 4.7 K. The red line indicates the spectrum of the OPO laser. (c) Optical setup used for PE experiment. (PBS: polarizing beam splitter, BS: half beam splitter, $\lambda/2$: half wave plate, $\lambda/4$: quarter wave plate.) (d) GVD curve of the glass rod based on literature values at room temperature. (e) Experimental result of GDD with different glass rod length. The red solid line represents a fit to the experimental data.}
\end{figure}

As a method for reading out signals stored in the QD ensemble, we have combined the revival of silenced echo (ROSE) scheme~\cite{Damon_2011} with the ARP technique. In this protocol, two rephasing pulses are used to recover macroscopic coherence, both realized as chirped pulses to induce the ARP effect, as shown in the pulse sequence in Fig. 1(a). The temporal relationship between the pulses and the PE are given by
\begin{equation}
    t_{\rm PE} = t_{\rm signal} + 2(t_2 - t_1) = t_{\rm signal} + 2 t_{21},
\end{equation}
where \(t_{\rm signal}\), \(t_1\), \(t_2\), and \(t_{\rm PE}\) denote the arrival times of the signal, rephasing 1, rephasing 2, and PE pulses, respectively, and \(t_{21} = t_2 - t_1\) is the interval between the two rephasing pulses. By adjusting \(t_{21}\), the storage time \(t_{\rm PE}\) can be arbitrarily controlled. For simplicity, we set \(t_{\rm signal} = 0\). According to Eq.~(1), the PE signal occurs only when the timing satisfies \(t_2 > 2 t_1\). The PE signal is obtained as the sum of the Bloch vectors over the ensemble's detuning. Therefore, achieving robust quantum control across the inhomogeneous ensemble is crucial for generating a high efficiency PE. In this context, ARP is considered a powerful tool to provide such robust control.

In this work, we first analyze the behavior of ARP in a two-level system ensemble under the assumption of resonant excitation of the exciton in InAs QDs.
We consider a linearly chirped Gaussian pulse, whose electric field is given by
\begin{equation}
    E(t) = E_0 \exp\left(-\frac{t^2}{2\tau^2}-i\omega_0 t - i \alpha \frac{t^2}{2}\right),
\end{equation}
\begin{equation}
    E(\omega) = E_0' \exp\left[-\frac{(\omega-\omega_0)^2}{2}\tau_0^2 +i \alpha' \frac{(\omega-\omega_0)^2}{2}\right],
\end{equation}
where $E_0,\ E_0'$ is the peak amplitude, $\omega_0$ is the center frequency, $\tau$ is the pulse width, related to the full width at half-maximum (FWHM) by $2\sqrt{\rm ln 2}\,\tau \equiv \tau_{\rm FWHM}$, $\alpha$ is the linear temporal chirp and $\alpha'$ is the linear spectral chirp \cite{Malinovsky01}. This $\alpha'$ is equivalent to the group delay dispersion (GDD).
The pulse width $\tau$ and the transform-limited (TFL) pulse width $\tau_0$ are related by
\begin{equation}
    \tau = \tau_0\sqrt{1 + \frac{\alpha'^2}{\tau_0^4}}.
\end{equation}
The time evolution of the quantum state $\bm{R}$ of each QD is described by the Bloch equation
\begin{equation}
    \frac{\partial}{\partial t} \bm{R} = \bm{\Omega} \times \bm{R},
\end{equation}
where $\bm{\Omega}$ is the torque vector. 
The torque vector $\bm{\Omega}$ and the pulse area $\Theta$ of the Bloch vector are expressed as
\begin{equation}
    \bm{\Omega} =
    \begin{pmatrix}
        \frac{2\mu E}{\hbar} \\
        0 \\
        \Delta - \alpha t
    \end{pmatrix},
\end{equation}
\begin{equation}
    \Theta = \int \sqrt{\left(\frac{2\mu E}{\hbar}\right)^2 + \Delta^2} \, dt,
\end{equation}
where $\mu$ is the dipole moment of the QD ensemble, $\Delta$ denotes the detuning of each QD from the center frequency of the rephasing pulse, and $\alpha$ represents the linear temporal chirp. 

Since $\alpha=0$ when the rephasing pulse is TFL, it can be seen that both the torque vector and the pulse area are affected by the detuning and the inhomogeneous distribution of the electric field. In the actual system, detuning $\Delta$ from the center frequency of the rephasing pulse becomes a serious issue because the inhomogeneous distribution of the resonance frequency of the QD ensemble extends to the THz range. In addition, the intensity distribution of the laser cross section generally follows a Gaussian distribution. Therefore, the electric field amplitude is different depending on the position of the QD on the sample, and there is also an inhomogeneity of the electric field. For these reasons, the storage efficiency of broadband quantum memory has been significantly low.

On the other hand, if the frequency variation of the chirped pulse is sufficiently large, the torque vector gradually moves slowly from one pole to the other, and every Bloch vector adiabatically follows this motion. As a result, ARP enables robust control over variations in $\Delta$ and electric field amplitude $E$. Applying this technique to QD ensembles is expected to address some of the challenges currently faced with these systems.
To compare the experimental results and theoretical values, we have performed numerical simulations of both population inversion and PE protocol. The Hamiltonian when a chirped pulse is applied can be shown as follows.
\begin{equation}
    H(t)=
    \begin{pmatrix}
    \frac{\hbar\left(\Delta-\alpha t\right)}{2} & -\mu E \\
    -\mu E & -\frac{\hbar\left(\Delta-\alpha t\right)}{2} \\
    \end{pmatrix}.
\end{equation}
In our simulation, we used the density matrix obtained from the Lindblad equation 
\begin{equation}
    \begin{split}
    \frac{\partial}{\partial t}\rho_{\Delta, E_0} &= -\frac{i}{\hbar} [H, \rho_{\Delta, E_0}] \\
    &+ \sum_k \left( C_k \rho_{\Delta, E_0} C_k^\dagger - \frac{1}{2} \left\{ C_k^\dagger C_k, \rho_{\Delta, E_0} \right\} \right)
    \end{split}
\end{equation}
where the collapse operators \(C_k\) are defined using the relaxation constants
\begin{equation}
\Gamma_1 = \frac{1}{T_1}, \quad \Gamma_2 = \frac{1}{T_2}, \quad \Gamma_\phi = \Gamma_2 - \frac{\Gamma_1}{2}
\end{equation}
as follows:
\begin{equation}
C_1 = \sqrt{\Gamma_1} \sigma_-, \quad C_2 = \sqrt{\frac{\Gamma_\phi}{2}} \sigma_z.
\end{equation}
$T_1$ and $T_2$ are the population relaxation time and the dephasing time, respectively.
We then calculated the time evolution of the Hamiltonian using the Python module QuTiP \cite{JOHANSSON20121760}, with the simulation parameters summarized in Table 1.

\begin{table}[htbp]
 \caption{Parameters used in the simulation of single $\pi$ rotation and PE.}
 \centering
  \begin{tabular}{lc}
   \hline
    Parameter & Value \\
   \hline
    Rephasing pulse diameter & $100\ \mu\rm m$ \\
    Signal pulse diameter & $200\ \mu\rm m$ \\
    Transition dipole moment $\mu$ & $57\ \rm Debye$ \\
    Population relaxation time $T_1$ & $1\ \rm ns$ \\
    Dephasing time $T_2$ & $600\ \rm ps$ \\
    Initial condition (single $\pi$ rotation) & $\ket{g}$ \\
    Initial condition (photon echo) & $\frac{1}{\sqrt{2}}(\ket{g}+\ket{e})$ \\
   \hline
  \end{tabular}
\end{table}

The QD ensemble used in our study has a density of approximately \(10^{12}~\mathrm{cm^{-2}}\), and the excitation laser beam has a diameter of about \(100~\mu\mathrm{m}\). This means that more than \(10^7\) QDs are simultaneously irradiated, which justifies treating the distributions over detuning and spatial position as continuous functions.
We approximate the inhomogeneous broadening of the QD ensemble by a Gaussian distribution in detuning, $\Delta$, and model the spatial variation of the electric field using a two-dimensional Gaussian beam profile. Here, $w_0$ is the beam radius, and $r$ is the radial distance from the center of the beam. The electric field amplitude at a radial position $r$ is expressed as
\begin{equation}
    E(r) = E_0 \exp\left(-\frac{r^2}{w_0^2}\right).
\end{equation}
The spectral weight function $A(\Delta)$ is modeled as a normalized Gaussian distribution centered at zero detuning:
\begin{equation}
    A(\Delta) = \frac{1}{\sigma \sqrt{2\pi}} \exp\left(-\frac{\Delta^2}{2\sigma^2}\right),
    \quad
    \sigma = \frac{\Delta_\mathrm{FWHM}}{2 \sqrt{2 \ln 2}},
\end{equation}
where $\Delta_\mathrm{FWHM}$ is FWHM of the spectral distribution. 
The total expectation value is then obtained by integrating over all spectral detunings $\Delta$ and radial positions $r$ in the transverse plane:
\begin{equation}
    \rho_{\rm total}(t) = \int_{-\infty}^{\infty} \int_0^\infty 2 \pi r \, A(\Delta) \, \rho_{\Delta, E(r)}(t) \, dr \, d\Delta.
\end{equation}
Fig.~1(b) shows the population inversion simulation results with different detuning values \( \Delta = \pm 1, \pm 0.6, \pm 0.4~\mathrm{THz} \) and pulse areas (electric field) \( E = 0.6\pi,\ \pi,\ 2\pi \). These $\Delta$ values reflect the typical spectral inhomogeneous broadening in self-assembled InAs QD ensembles. Compared to the case with the TFL pulse, almost all Bloch vectors are driven to the excited state when a chirped pulse is applied. This result confirms that the ARP technique is effective for achieving robust control even in materials with THz-range inhomogeneous broadening. For the full motion of the Bloch vectors, see Supplementary Video 1.

\begin{figure*}[t]
\includegraphics[width=\linewidth]{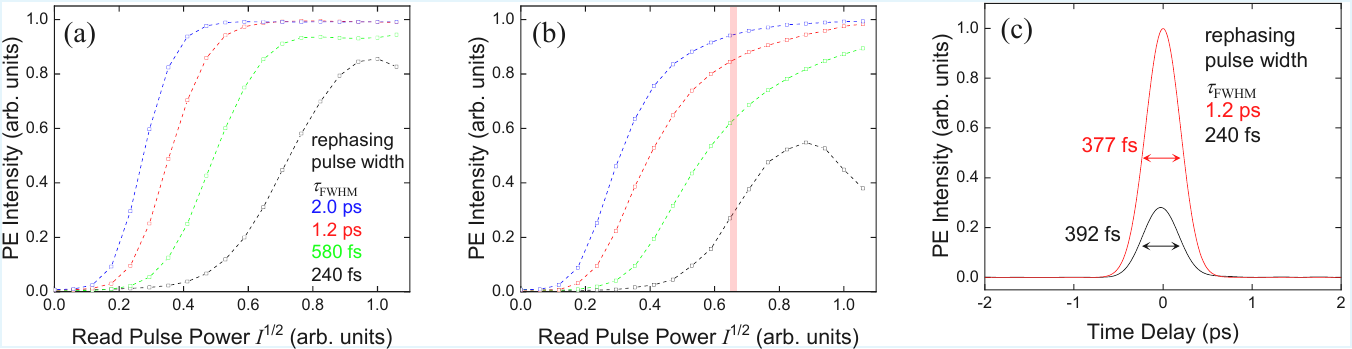}
\caption{Simulation results for pulse area dependence of PE intensity: (a) with a flat-top and (b) with a Gaussian electric field profile. The red area indicates the pulse area in the PE waveform measurement experiment. PE waveforms corresponding to the red area $\sqrt{I}=0.65$ in (b) are shown in (c).}
\end{figure*}

\subsection{\label{sec:level2}ARP-enhanced photon echo}
While the principle of ARP is similar to that of conventional population inversion, a key distinction in PE scheme is that the initial quantum state is a coherent superposition, and two chirped pulses are applied to achieve rephasing. In our PE protocol (Fig.~1(a)), a weak signal pulse (with wave vector \( \bm{k}_{\rm signal} \)) is first applied to the QD ensemble, generating an excitonic superposition state. Subsequently, two rephasing pulses (denoted as rephasing~1 and rephasing~2, or \( \bm{k}_1 \) and \( \bm{k}_2 \), respectively) are used to retrieve the macroscopic coherence as a photon echo signal with wave vector \( \bm{k}_{\rm PE} \) at $t_{\rm PE}=t_{\rm signal} + 2 t_{21}$.
The directional relationships between the pulses and the resulting PE are given by
\begin{equation}
    \bm{k}_{\rm PE} = \bm{k}_{\rm signal} + 2(\bm{k}_2 - \bm{k}_1).
\end{equation}
In our PE simulations, we employed the time-dependent Hamiltonian described in Eq.~(8), which explicitly includes all three optical pulses: the signal pulse, the first rephasing pulse, and the second rephasing pulse. The chirp parameter \(\alpha\) was adjusted for each pulse to match the experimentally measured pulse durations. During intervals when no pulse was present, the system evolved under the free precession Hamiltonian given by
\begin{equation}
    H_{\rm free} =
    \begin{pmatrix}
    \frac{\hbar \Delta}{2} & 0 \\
    0 & -\frac{\hbar \Delta}{2}
    \end{pmatrix}.
\end{equation}
The total Hamiltonian \( H_{\rm total}(t) \) was defined as
\begin{equation}
H_{\rm total}(t) =
\begin{cases}
H_{\rm signal}, & -3\tau_{\rm signal} \leq t < 3\tau_{\rm signal}, \\
H_{\rm{reph. 1}}, & t_1 - 3\tau_1 \leq t < t_1 + 3\tau_1, \\
H_{\rm{reph. 2}}, & t_2 - 3\tau_2 \leq t < t_2 + 3\tau_2, \\
H_{\rm free}, & \text{otherwise},
\end{cases}
\end{equation}
where \( H_{\rm signal} \), \( H_{\rm{reph. 1}} \), and \( H_{\rm{reph. 2}} \) are corresponding to Hamiltonian for the signal, rephasing 1, and rephasing 2, respectively, defined by Eq.(8). Each Hamiltonian uses a different value of the chirp parameter \(\alpha\) to reflect the properties of the respective pulses. The parameters \( \tau_{\rm signal} \), \( \tau_1 \), and \( \tau_2 \) represent the temporal standard deviations of each pulse in Eq. (2), and the cutoff of \( \pm 3\tau \) ensures numerical accuracy while maintaining physical relevance. The PE signal was finally obtained by solving the Lindblad equation given in Eq. (9).

Figure~1(c) illustrates the final states of the PE protocol for Bloch vectors with different detunings \(\Delta = \pm 1, \pm 0.6, \pm 0.4\, \mathrm{THz}\) and pulse areas \(\Theta = 0.6 \pi, \pi, 2\pi\). For TFL rephasing pulses, the final states show significant dispersion, resulting in weaker macroscopic coherence and thus a reduced PE signal. In contrast, with 1.2 ps chirped pulses (ARP), the Bloch vectors are nearly aligned, restoring macroscopic coherence. This demonstrates that even starting from a superposition state, the application of two chirped rephasing pulses enables robust control against variations in THz-scale detuning \(\Delta\) and field amplitude \(E\).
The full Bloch vector dynamics are provided in Supplementary Video~2.

\subsection{\label{sec:level2}Optical setup and InAs QD ensemble}

We utilized a self-assembled InAs QD ensemble with a resonator structure. Fig. 2(a) presents the cross sectional SEM image of our sample. This sample comprises 20 pairs of semiconductor (InP/InGaAlAs) distributed Bragg reflector (DBR) on an InP(311) substrate, 50 layers of InAs QD ensemble stacked using a strain-compensation method, and three pairs of dielectric DBR consisting of $\rm{SiO_2/TiO_2}$ on top of this structure\cite{Akahane02, Akahane11}. This strain-compensation technique enables the center resonant frequency into the telecom band. The areal density of QDs was approximately \(2.9 \times 10^{12}\,\mathrm{cm^{-2}}\), with average dot dimensions of 65\,nm along the \([ \bar{2}33 ]\) crystallographic direction and 50\,nm along \([01\bar{1}]\). By employing linearly polarized light, selective excitation along a specific axis is achieved, effectively enabling the system to be treated as a two-level artificial atom. The semiconductor DBR and dielectric DBR form a low-Q ($\sim$100) Fabry-Pérot resonator. This configuration enhances the overall electric field acting on the QD ensemble while maintaining THz-scale inhomogeneous broadening. Fig. 2(b) shows photoluminescence measurements from our QD ensemble sample at 4.7 K, indicating that the inhomogeneous broadening was 3.1~THz. The polarization directions of these pulses were aligned along the $[\bar{2}33]$ axis of the QD ensemble to selectively excite either of the exciton states and maximize the PE signal intensity.

The pulse sequence for PE using ARP is shown in Fig. 1(a), and an overview of our optical setup is depicted in Fig. 2(c). We employed an optical parametric oscillator (OPO) with a pulse width of 240 fs and a wavelength of 1,542\,nm, which corresponds to the central resonance wavelength of the QD ensemble. Its spectral width was 3.2 THz, so the pulse width was slightly broader than the TFL pulse width ($\rm{GDD}=1.0 \times 10^{4}\ \rm fs^2$). The OPO was pumped by a mode-locked Ti:Sapphire laser (76.4\,MHz, 820\,nm). The QD ensemble was cooled to 8.3 K in a cryostat for PE experiment. The two rephasing pulses were aligned coaxially ($\bm{k}_1 = \bm{k}_2$), resulting in $\bm{k}_{\rm PE}$ being equal to $\bm{k}_{\rm signal}$ and the PE signal being generated in the same direction as the signal pulse.

Instead of using a conventional Offner pulse stretcher based on a diffraction grating, we generated positively chirped pulses by varying the pulse width $\tau_{\rm FWHM}$ from 240 fs to 1.20 ps using glass rods (OHARA, S-NPH3, length $L=20\sim60$ cm). This approach aims to minimize power losses associated with diffraction gratings. Positive chirp was deliberately chosen because previous studies have shown that positively chirped pulses are more efficient for ARP than negatively chirped pulses \cite{Kaldewey17, Wei14}. The literature value of refractive index of the glass is 1.89, and the group velocity dispersion (GVD) is $76.6,\mathrm{fs^2/mm}$ at a wavelength of 1,542 nm (Fig. 2(d)) \cite{Mikhail24}. Fig. 2(e) shows the experimental values of GDD on the glass rod length, from which we obtained a GVD value of \(77.7\ \mathrm{fs^2/mm}\), which is in good agreement with the literature value. The PE signal was measured via heterodyne detection using a balanced photodetector, and the temporal waveform was obtained by scanning the reference delay stage.

\begin{figure*}[t]
\includegraphics[width=16cm]{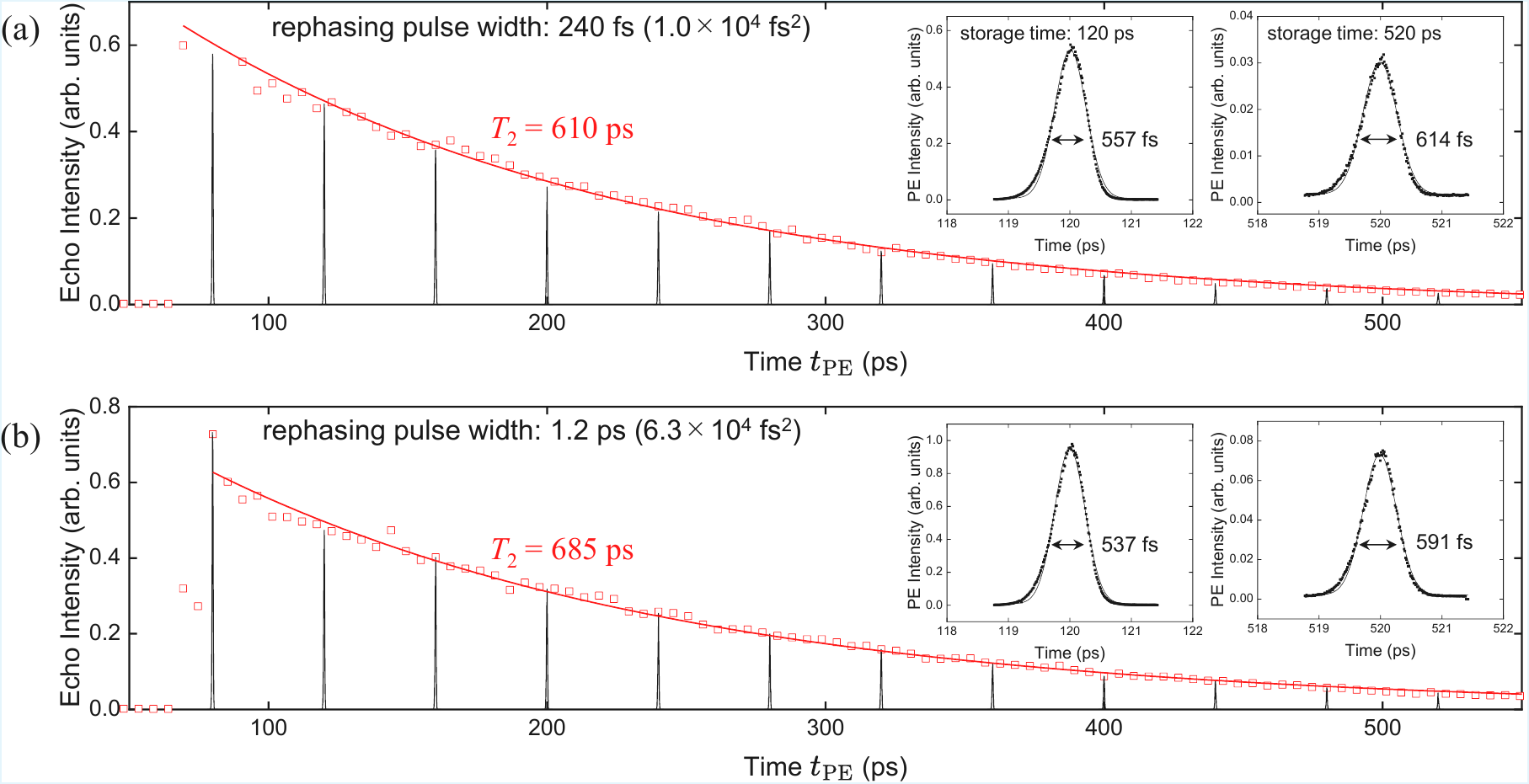}
\caption{Storage time $t_{\rm PE}$ dependence of PE intensity (a) without and (b) with ARP. Red squares denote peak PE intensity, red solid lines are exponential fits, and black solid lines are PE waveforms. Insets show expanded PE traces at 120 and 520 ps.}
\end{figure*}

\section{\label{sec:level1}Results and discussion}
\subsection{\label{sec:level2}Simulation results of ARP-enhanced photon echo}

Figure~3(a) shows the dependence of the photon echo (PE) intensity on the square root of the rephasing pulse intensity, $\sqrt{I}$, for spatial flat-top pulses. 
We simulated four rephasing pulses with different group delay dispersions (GDD):$1.0 \times 10^{4}$, $3.0 \times 10^{4}$, $6.3 \times 10^{4}$, and $1.0\times 10^5~\mathrm{fs^2}$, corresponding to pulse widths $\tau_{\rm FWHM}$ of 240~fs, 580~fs, 1.2~ps, and 2.0~ps, respectively. For pulses at the same intensity $I$, the effective pulse area $\Theta$ increases with pulse width as $\Theta \propto \sqrt{\tau/\tau_0}$, implying that pulse stretching enhances $\Theta$ even at fixed laser power. As seen in Fig. 3(a), clear Rabi oscillations appear for near-TFL pulses, but they vanish with increasing chirp, resulting in a saturation of the PE intensity. These results demonstrate that ARP suppresses pulse-area dependence, enabling robust rephasing in the femtosecond PE process. 
For example, with $\tau_{\rm FWHM} = 580$~fs, the maximum PE intensity reaches 93\% of the ideal value, whereas for pulses longer than 1.2~ps it recovers the full intensity. 
This behavior is consistent with the fact that the frequency sweep range, $2 \alpha \tau$, reaches the inhomogeneous width of our QD ensemble, $3.1~\mathrm{THz}$, for GDD $\alpha' \ge 4.0 \times 10^{4}~\mathrm{fs^2}$, with a rephasing pulse width of $\tau_\mathrm{FWHM} = 670~\mathrm{fs}$. Notably, ARP provides significant enhancement at relatively low intensities, where conventional TFL rephasing fails.

Figure 3(b) presents analogous simulations for rephasing pulses with a Gaussian spatial profile, chosen to match experimental beam sizes (100 $\mu$m for the signal and 200 $\mu$m for the rephasing pulse). Compared to the flat-top case, the PE intensity is reduced by spatial averaging, yet robust rephasing re-emerges at higher intensities. The corresponding temporal waveforms are shown in Fig. 3(c). At $\sqrt{I}=0.65$ (red region in Fig. 3(b)), the simulated PE widths are 392 fs without ARP and 377 fs with ARP, indicating that ARP enhances efficiency without degrading temporal bandwidth. The shortening of the PE pulse with ARP is due to contributions from QDs with large $\Delta$, which broaden the PE spectrum. The calculated signal enhancement factor is 3.56.
Taken together, these simulations predict that ARP not only mitigates sensitivity to pulse area and spatial inhomogeneity but also preserves sub-picosecond bandwidth, providing a viable route to efficient PE generation in broadband, inhomogeneous systems.

\subsection{\label{sec:level2}Experimental results and comparison with theoretical model}

We next tested these predictions experimentally, measuring the storage-time and pulse-area dependence of PE signals in a QD ensemble. In this experiment, we fixed the position of rephasing 1 at $t_1 = 35$ ps and varied rephasing 2 to alter the storage time.

\begin{figure*}[t]
\centering
\includegraphics[width=13cm]{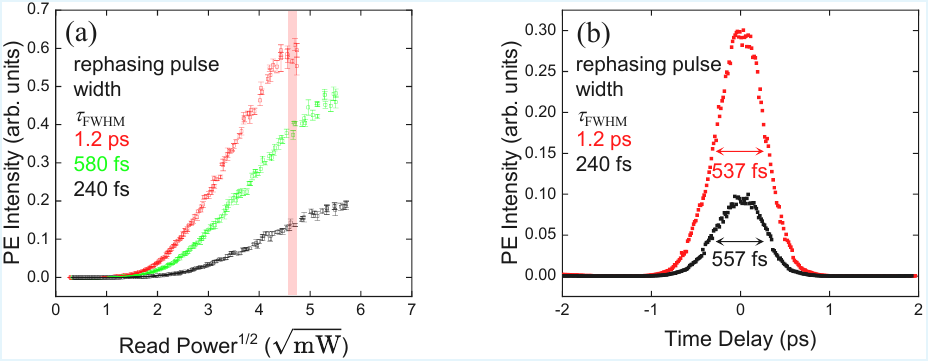}
\caption{(a) Experimental results of PE pulse area dependence with different chirp amounts. (b) PE waveforms corresponding to the red areas in (a). Experimental waveforms were obtained using heterodyne detection, resulting in cross-correlated waveforms. The pulse width values shown are estimates of the original pulse width, derived from deconvolution of the waveforms after fitting with a Gaussian function.}
\end{figure*}

Figure 4 compares the storage-time dependence of PE intensity using 240 fs near-TFL rephasing pulses ($1.0\times10^{4}$ fs$^2$) and 1.2 ps chirped pulses ($6.3\times10^{4}$ fs$^2$), both with 22.4 mW average power. The absence of PE signals below 70 ps storage time is due to the time range in which PE does not occur, as determined by $t_{\rm{1}}$ value. The exponential fits $\exp\left(-4t/T_2\right)$ yield dephasing times $T_2$ of 610 ps and 685 ps, respectively. The modest extension of $T_2\ (610\rightarrow685\ \rm ps)$ primarily reflects reduced excitation-induced dephasing due to chirp-induced peak power reduction, while ARP ensures robust rephasing across detuning variations. For weak excitation, $T_2$ extended to $\sim$1.5 ns. The application of ARP enables robust control even at lower peak powers, so the coherence decay is less pronounced than with high peak power TFL pulses. Insets show representative PE waveforms, from which actual widths were estimated via the original OPO pulse width. At $t_{\rm PE}=120$ ps, PE widths were 557 fs (TFL) and 537 fs (ARP), while at $t_{\rm PE}=520$ ps they broadened to 614 fs and 591 fs, respectively. This PE pulse width shortening with ARP reflects robust control over frequency detuning, while the broadening at longer storage times originates from ensemble decoherence. From extrapolation to $t_{\rm PE}=0$, the maximum PE generation efficiency was $4.2 \times 10^{-3}$\% without ARP and $1.3 \times 10^{-2}$\% with ARP. The efficiency is still limited due to the small absorption of our InAs QDs sample. Improvements could be achieved by increasing absorption via waveguide structures or by enlarging the beam diameter to address more QDs. These approaches would increase the overall PE efficiency while maintaining the enhancement effect of ARP, suggesting that ARP will continue to be a powerful tool in future experiments.

The pulse-area dependence of PE intensity is summarized in Fig. 5(a). The results closely match the simulation trends in Fig. 3(b), confirming that ARP indeed governs the rephasing dynamics. Temporal waveforms at equal pulse power (22.4 mW, red regions) are shown in Fig. 5(b), demonstrating a 3.2-fold enhancement in PE intensity with ARP. The measured widths were 557 fs without ARP and 537 fs with ARP, consistent with preserved ultrafast bandwidth. This enhancement factor is comparable to those reported in rare-earth ion-doped crystals, underscoring the robustness of ARP in highly inhomogeneous semiconductor ensembles. The slight discrepancy between experiment and simulation is attributed to residual chirp in the echo and spectral filtering by the resonator.

Our combined simulations and experiments demonstrate that ARP enables reliable PE rephasing in broadband, THz-scale inhomogeneous ensembles. Quantitatively, ARP increased the PE efficiency by a factor of 3.2 and preserved femtosecond temporal widths. These results establish ARP as a practical strategy for coherent control of QD ensembles, opening new opportunities for broadband quantum memory and ultrafast quantum information processing compatible with telecommunication wavelengths.

\section{\label{sec:level1}Summary}

We have demonstrated, for the first time, the application of ARP to a highly inhomogeneous quantum system in the THz spectral regime: a self-assembled InAs QD ensemble integrated with a resonator. By employing chirped optical pulses satisfying adiabatic conditions across the ensemble’s ultrabroad linewidth, we achieved robust quantum control despite spectral detuning and field inhomogeneity. Experimentally, this resulted in up to a 3.2-fold enhancement in PE intensity compared to transform-limited pulses, in agreement with numerical simulations. Despite the THz-scale inhomogeneous broadening, we observe this level of enhancement comparable to that in rare-earth ion–doped crystals. Furthermore, the PE efficiency is expected to improve with optimized QD samples, and the combination of THz-bandwidth excitation with ARP enhancement demonstrates the significant potential of this approach.

Simulations confirmed that, even with chirped pulses for ARP, the PE temporal width remains in femtosecond regime. This work provides the first demonstration of ARP at femtosecond timescales, showing that ultrafast pulses can be coherently rephased without broadening. Such control enables manipulation of collective coherence and sub-picosecond pulse storage in the telecommunication band, paving the way for ultrafast quantum communication and broadband photonic quantum technologies.

Future work will focus on optimizing the excitation beam profile and mode matching between the optical field and the QD ensemble. In particular, aligning the resonator’s resonance with the QD absorption peak near 1,520 nm and employing flat-top beams with uniform intensity are expected to further enhance PE signal generation, advancing the realization of scalable, high-speed quantum memory devices and ultrafast nonlinear optical applications.

\begin{acknowledgments}
This work was supported by the Grant-in-Aid for JSPS Fellows (Grant No. JP23KJ1916), JST CREST (Grant No. JPMJCR24A5), the Center for Spintronics Research Network, and the Program for the Advancement of Next Generation Research Projects at Keio University. The QD samples were fabricated at the Advanced ICT Laboratory, NICT.
\end{acknowledgments}

\bibliography{apssamp}

\end{document}